\documentclass[conference]{IEEEtran}
\IEEEoverridecommandlockouts
\usepackage{cite}
\usepackage{amsmath,amssymb,amsfonts}
\usepackage{algorithmic}
\usepackage{graphicx}
\usepackage{textcomp}
\usepackage{xcolor}
\usepackage{siunitx}

\def\BibTeX{{\rm B\kern-.05em{\sc i\kern-.025em b}\kern-.08em
    T\kern-.1667em\lower.7ex\hbox{E}\kern-.125emX}}
\begin{document}

\title{Efficient Reinforcement Learning On Passive RRAM Crossbar Array}

\author{\IEEEauthorblockN{Arjun Tyagi}
\IEEEauthorblockA{\textit{Department of Electrical and Electronics Engineering} \\
\textit{Birla Institute of Technology and Science Pilani}\\
Goa, India \\
f20190579g@alumni.bits-pilani.ac.in}
\and
\IEEEauthorblockN{Shubham Sahay}
\IEEEauthorblockA{\textit{Department of Electrical Engineering} \\
\textit{Indian Institute of Technology Kanpur}\\
Kanpur, India\\
ssahay@iitk.ac.in}
}

\maketitle

\begin{abstract}
The unprecedented growth in the field of machine learning has led to the development of deep neuromorphic networks trained on labelled dataset with capability to mimic or even exceed human capabilities. However, for applications involving continuous decision making in unknown environments, such as rovers for space exploration, robots, unmanned aerial vehicles, etc., explicit supervision and generation of labelled data set is extremely difficult and expensive. Reinforcement learning (RL) allows the agents to take decisions without any (human/external) supervision or training on labelled dataset. However, the conventional implementations of RL on advanced digital CPUs/GPUs incur a significantly large power dissipation owing to their inherent von-Neumann architecture. Although crossbar arrays of emerging non-volatile memories such as resistive (R)RAMs with their innate capability to perform energy-efficient \textit{in situ} multiply-accumulate operation appear promising for Q-learning-based RL implementations, their limited endurance restricts their application in practical RL systems with overwhelming weight updates. To address this issue and realize the true potential of RRAM-based RL implementations, in this work, for the first time, we perform an algorithm-hardware co-design and propose a novel implementation of Monte Carlo (MC) RL algorithm on passive RRAM crossbar array. We analyse the performance of the proposed MC RL implementation on the classical cart-pole problem and demonstrate that it not only outperforms the prior digital and active 1-Transistor-1-RRAM (1T1R)-based implementations by more than five orders of magnitude in terms of area but is also robust against the spatial and temporal variations and endurance failure of RRAMs.
\end{abstract}


\section{Introduction}
Reinforcement learning (RL) \cite{sutton2018reinforcement} has been the driving force behind major breakthroughs in the field of autonomous driving, industrial automation, robotics, etc. Moreover, the recent developements in deep neural networks \cite{mnih2013playing} have further advanced reinforcement learning, leading to achievements like AlphaGo (first computer program to defeat a Go world champion). However, the first AlphaGo was trained on 1920 central processing units (CPUs) and 280 graphic processing units (GPUs), consuming a peak power of 0.5 \si{\mega\watt} \cite{1T1R_RL}. This large energy consumption is primarily attributed to the frequent back-and-forth data movement between the main memory and the von-Neumann processing engines such as CPUs and GPUs (at least four orders of magnitude more than the energy required to process the data itself \cite{haj2018not}).

\begin{figure}
    \centering
    \includegraphics[width=0.5\textwidth]{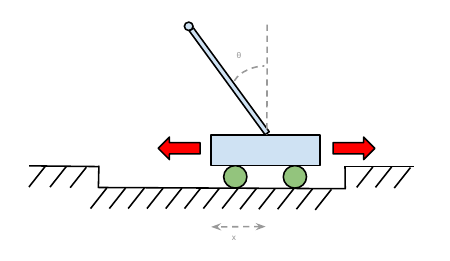}
    \caption{Schematic illustration of the Cart-Pole environment. The cart is free to move in the one-dimensional bounded track.}
    \label{fig:cart_pole}
\end{figure}

Therefore, for efficient hardware implementation of RL algorithms, several approaches have been explored including processing-in-memory (PIM) architecture, which shows an inherent ability to perform the fundamental operations such as vector-by-matrix multiplication at the storage location itself without energy draining data transfers, with the help of physical laws such as Ohm's law and Kirchoff's law \cite{1T1R_RL}. Recently, ferroelectric tunnel junction (FTJ)-based PIM implementations of reinforcement learning algorithm were demonstrated in \cite{ota2019performance} and \cite{berdan2019memory} on the standard Cart-Pole control problem (Fig. \ref{fig:cart_pole}). However, the large write voltage and the poor dynamic range (ON/OFF ratio) of the FTJs limit their widespread applications. Considering the low program/read voltages, high scalability, multi-level capability, and CMOS-compatibility of the resistive (R)RAMs \cite{RRAMarray1,RRAMarray2,RRAMarray3,RRAMarray4,RRAMarray5,RRAMarray6}, reinforcement learning on active 1T-1R crossbar array was experimentally demonstrated in \cite{1T1R_RL} and bench marked using the Cart-Pole control problem. However, the active 1T-1R crossbar arrays exhibit a large area overhead since they utilize a selector to minimize the sneak path leakage current and efficiently tune the conductance-state of the RRAMs. Recently, area-efficent passive RRAM crossbar arrays with optimized stack showing a high yield, low sneak path leakage current, high weight precision upto 6 bits were demonstrated experimentally \cite{RRAMarray1}. Therefore, it becomes imperative to explore the potential of passive RRAM crossbar arrays for efficient implementation of RL algorithms. 

Moreover, most of the prior hardware implementations have focused on neural network-based RL algorithms such as deep-Q learning which require exhaustive switching of the memory devices eventually pushing them towards their endurance failure limits during \textit{in situ} training. Since efficient training of RL agents is based on the rewards accumulated after taking continuous actions, most RL algorithms involve a large number of weight updates which may exceed the endurance limit of the practical memory devices. This may limit the size of the network and the lifetime of the agent restricting the utility of the PIM-based RL accelerators to simple problems. Therefore, there is an urgent need to explore alternate hardware friendly RL algorithms. 

Monte Carlo learning is a class of RL algorithm which differs from other RL algorithms in terms of weights update process. Instead of updating the weights after each action of the agent, Monte Carlo learning updates the weights after the completion of an episode. This leads to a significant reduction in the number of weight updates as compared to other RL algorithms like deep-Q learning, SARSA, etc. Considering the efficacy of the Monte Carlo learning and the passive RRAM crossbar arrays, in this work, for the first time, we perform an algorithm-hardware co-design and explore the hardware implementation of Monte Carlo learning on passive RRAM crossbar array considering the practical artifacts such as endurance failure, device-to-device variation, etc. The \textit{in situ} training is carried out on $12\times24$ passive RRAM crossbar array which is partitioned into two sections of $6\times24$ each. With the aid of an experimentally calibrated compact model for the passive RRAM crossbar array, we show that Monte Carlo learning can be applied to classic RL environments such as Cart-Pole without pushing the RRAM devices close to their endurance limits. Our comprehensive analysis indicates that the proposed implementation of Monte Carlo learning on passive RRAM crossbar array achieves an area reduction of $\sim1.18\times10^5$ over active 1T-1R crossbar array while exhibiting a similar performance.

\begin{figure}
    \centering
    \includegraphics[width=0.5\textwidth]{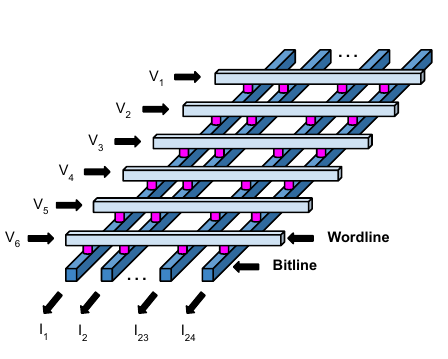}
    \caption{3-D view of a passive RRAM crossbar array.}
    \label{fig:passive_RRAM}
\end{figure}

\section{Reinforcement Learning}
RL algorithms rely on the interaction of an agent with the environment through actions and rewards \cite{RLIntro}. An RL agent can be defined using two approaches, namely policy-based agent and value-based agent \cite{RLIntro}. The aim of a policy-based agent is to find a function that maps each and every state to the best possible action to maximize the reward that the agent accumulates. The policy can either be a deterministic policy wherein each state has a certain action associated with it or it can be a stochastic policy where a probability distribution is used to map each action to every state.

On the other hand, the aim of a value-based agent is to represent the best possible action for each state. The value-based agent utilizes a value function V\textsubscript{$\pi$} that associates a value with each state-action pair. The value is the expected return (summation of rewards) from each state-action pair till termination of the episode. Therefore, the value function emphasizes more on long-term gains for a particular action rather than the short-term gains. The value function can be mathematically represented as: 

\begin{equation}
    v\textsubscript{$\pi$}(s) = E\textsubscript{$\pi$}(R\textsubscript{t}|s\textsubscript{t}=s) = E\textsubscript{$\pi$}(\sum^{\infty}_{k=0} \gamma^k r_{t+k+1}|s\textsubscript{t}=s)
    \label{value_function}
\end{equation}

where E\textsubscript{$\pi$} denotes the expected value given that the agent follows a policy $\pi$, R\textsubscript{t} and r\textsubscript{t} are return and reward at any time step t and $\gamma$ denotes the discount factor. 

\subsection{Monte Carlo Learning: First Visit}
One of the powerful yet simple RL algorithm is the Monte Carlo (MC) Learning. Due on a subtle difference in the methodology of updating the state-value, MC Learning is further distinguished into MC - First Visit and MC - Every Visit \cite{RLIntro}. MC - First Visit estimates the state-value v\textsubscript{$\pi$}(s) for a state \textit{s} under a policy $\pi$ by averaging the return from all the sampled episodes for that particular state \textit{s}. Furthermore, the presence of a termination state is necessary since the value function can calculate the return from each state only after an episode terminates. Therefore, MC - First Visit can only be applied to episodic problems. For a state \textit{s}, the state-value is updated at the end of each episode using:

\begin{equation}
    v\textsubscript{$\pi$}(s) \leftarrow v\textsubscript{$\pi$}(s) + \frac{1}{N(s)}(G\textsubscript{s} - v\textsubscript{$\pi$}(s))
    \label{inc_mean}
\end{equation}

where N(s) denotes the number of times state \textit{s} is visited during the course of training, G\textsubscript{s} represents the return from state \textit{s} till termination and v\textsubscript{$\pi$} is the current value for state \textit{s}.
As the agents is trained on more and more episodes, v\textsubscript{$\pi$} converges to the expected return from each state till termination.

\section{Simulation Framework}
We explored the implementation of Monte Carlo learning algorithm on passive RRAM crossbar array shown in fig. \ref{fig:passive_RRAM}. For proof-of-concept demonstration, we simulated the Cart-Pole environment using an agent trained on Monte Carlo learning and implemented the same environment on hardware using passive RRAM crossbar array. Monte Carlo learning on Cart-Pole environment was explored using two methodologies: (a) The digital approach where the equations governing the Monte Carlo learning were implemented as various MATLAB functions and (b) The hybrid approach where the integral computations were performed \textit{in situ} on passive RRAM crossbar array. For the hybrid approach, we also propose an algorithm which dictates the voltage pulses that are applied to each and every cell of the passive RRAM crossbar array. We encoded the state-value matrix used by the agent to take actions as the conductance states of passive RRAM crossbar array for \textit{in situ} training. For minimizing the number of programming cycles and energy consumption, the conductance range for each RRAM cell was restricted between 100 \si{\micro\siemens} and 300 \si{\micro\siemens} \cite{9651522}. For \textit{in situ} training of Monte Carlo learning, the ($12\times24$) passive RRAM crossbar array \cite{RRAMarray1} was partitioned into a $6\times24$ array (weight matrix) which stores the state-value that associates a value with each state in the environment and a $6\times24$ array (return matrix) which stores the return from each state visited by the agent during an episode. Furthermore, the partition ensures that weight matrix and return matrix on the passive RRAM crossbar array share the same bitlines. 

For encoding the weights as conductance states of RRAMs, a conductance-to-weight ratio of $2.5\times10^{-4}$ was selected for the weight matrix and return matrix. The conductance states for the $12\times24$ array was initialised to 200 \si{\micro\siemens} at the start of the \textit{in situ} training. At the end of each episode, the state-values in the weight matrix were updated according to the Manhattan rule \cite{ManhattanRule}, which is a hardware-friendly variation of the back-propagation algorithm, wherein the weights are updated based on the sign of the gradient instead of its exact value. Depending on whether the sign of $\Delta W(s,a) = R(s,a) - V(s,a)$ is positive or negative, the weights $V(s,a)$ in the weight matrix were potentiated or depressed. Furthermore, the weight updates were performed using one-shot voltage pulse of fixed duration of 100 \si{\nano\second} and fixed voltage (V\textsubscript{SET} = $0.8$V for potentiation and V\textsubscript{RESET} = $-0.8$V for depression. This results in a voltage V\textsubscript{RRAM} across the RRAM cell given by:

\begin{equation}
  V\textsubscript{RRAM}=\begin{cases}
    V\textsubscript{SET}, & \text{if $\Delta W(s,a) > 0$}.\\
    
    V\textsubscript{RESET}, & \text{if $\Delta W(s,a) < 0$}.\\
    
    0, & \text{otherwise}.
  \end{cases}
\end{equation}

As a result of the voltage pulse, the conductance of the RRAM cell gets updated according to the experimentally calibrated phenomological model for passive RRAM crossbar array \cite{RRAMarray4}:

\begin{equation}
\begin{gathered}
    G = G_0 + \Delta G_0 \\
    \Delta G_0 = D_m(G_0,V\textsubscript{RRAM},t_p) + D\textsubscript{d2d}(G_0,V\textsubscript{RRAM},t_p)
\end{gathered}
\end{equation}

where D\textsubscript{m} is the expected noise-free conductance change after application of voltage pulse and D\textsubscript{d2d} is the normally distributed device-to-device variation in passive RRAM crossbar array. 

To efficiently perform \textit{in situ} training of Monte Carlo learning on passive RRAM crossbar array using Manhattan learning rule, a novel 4-step algorithm that dictates the application of voltage pulses at the bitlines and wordlines of the weight matrix and return matrix at the end of each episode of the training is proposed:

\begin{figure}
    \centering
    \includegraphics[width=0.5\textwidth]{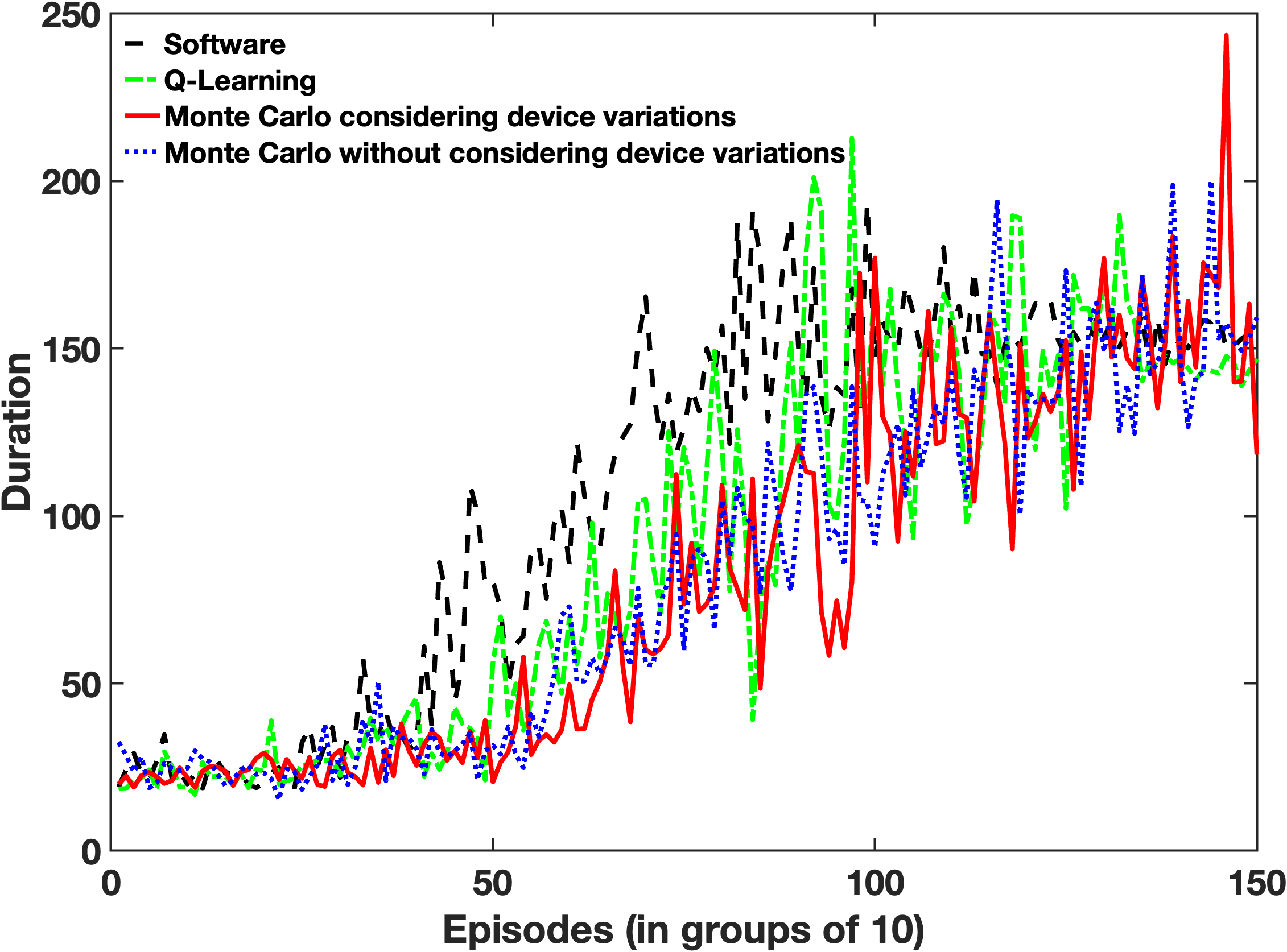}
    \caption{Reward accumulated by the agent during training using different Monte Carlo learning implementation for 1500 episodes.}
    \label{fig:mean_duration}
\end{figure}

\begin{enumerate}
    \item For each state-action pair that was performed in the episode, map the return for that state-action pair to the corresponding RRAM cell in the return matrix. For the remaining state-action pairs, their conductance state is kept same as their corresponding conductance state in the weight matrix.
    
    \item For i\textsuperscript{th} wordline in weight and return matrix, apply -V\textsubscript{READ} (usually equal to $|$V\textsubscript{SET}/2$|$) at the weight matrix wordline and V\textsubscript{READ} at the return matrix wordline while connecting all the bitlines to ground. The resulting current through each bitline is given by:
    
    \begin{equation} \label{current_eqn}
        I\textsubscript{j} = (G\textsubscript{i,j}\textsuperscript{R} - G\textsubscript{i,j}\textsuperscript{W}) \times V\textsubscript{READ}
    \end{equation}
    
    where i,j denotes the i\textsuperscript{th} wordline and j\textsuperscript{th} bitline and G\textsubscript{i,j}\textsuperscript{R}, G\textsubscript{i,j}\textsuperscript{W} denotes the conductance state of the return matrix and weight matrix, respectively. 
    
    \item Based on whether the current I\textsubscript{j} is positive or negative, the weight conductance state G\textsubscript{i,j}\textsuperscript{W} is potentiated or depressed.
\end{enumerate}

Furthermore, the weight updation step (step 3) for the weight matrix can be performed simultaneously for all RRAM cells in a single row \cite{haj2018not} to realize massive parallelism by connecting the i\textsuperscript{th} wordline to ground and applying either V\textsubscript{SET} or V\textsubscript{RESET} to the j\textsuperscript{th} bitline depending on whether the current I\textsubscript{j} in equation \ref{current_eqn} is positive or negative, respectively. This results in a change in the conductance state of all the RRAM cells in i\textsuperscript{th} row of the weight matrix. The proposed implementation of Monte Carlo learning on passive RRAM crossbar array was applied to the standard Cart-Pole problem similar to \cite{1T1R_RL} where the aim of the agent is to balance the pole mounted on a freely moving cart (within a bounded track) for a large duration by applying a force of fixed magnitude on the cart either in the left or right direction.

\section{Results \& Discussion}
For efficient benchmarking, we have carried out an extensive analysis of different figures of merit such as performance, energy consumption, number of programming cycles and area of the proposed MC learning implementation on passive RRAM crossbar array and compared them against the digital (software) and the prior RL implementation on the active 1T-1R array for the Cart-Pole problem.

\subsection{Performance}
The rewards accumulated by the agent during training for the digital (software) implementation and the proposed Monte Carlo learning implementation on passive RRAM crossbar array with and without the device variations and noise is shown in Fig. \ref{fig:mean_duration}. Even though the digital implementation of Monte Carlo learning converges faster, the proposed Monte Carlo learning implementation on passive RRAM crossbar array exhibits a similar performance. Moreover, the performance of the proposed Monte Carlo learning implementation of passive RRAM crossbar array considering the non-idealities such as noise and variability closely follows the behavior of the noise-free implementation. Therefore, the proposed Monte Carlo learning implementation on passive RRAM crossbar array is also resilient to the hardware artifacts such as spatial and temporal variations and noise.

\begin{figure}
    \centering
    \includegraphics[width=0.5\textwidth]{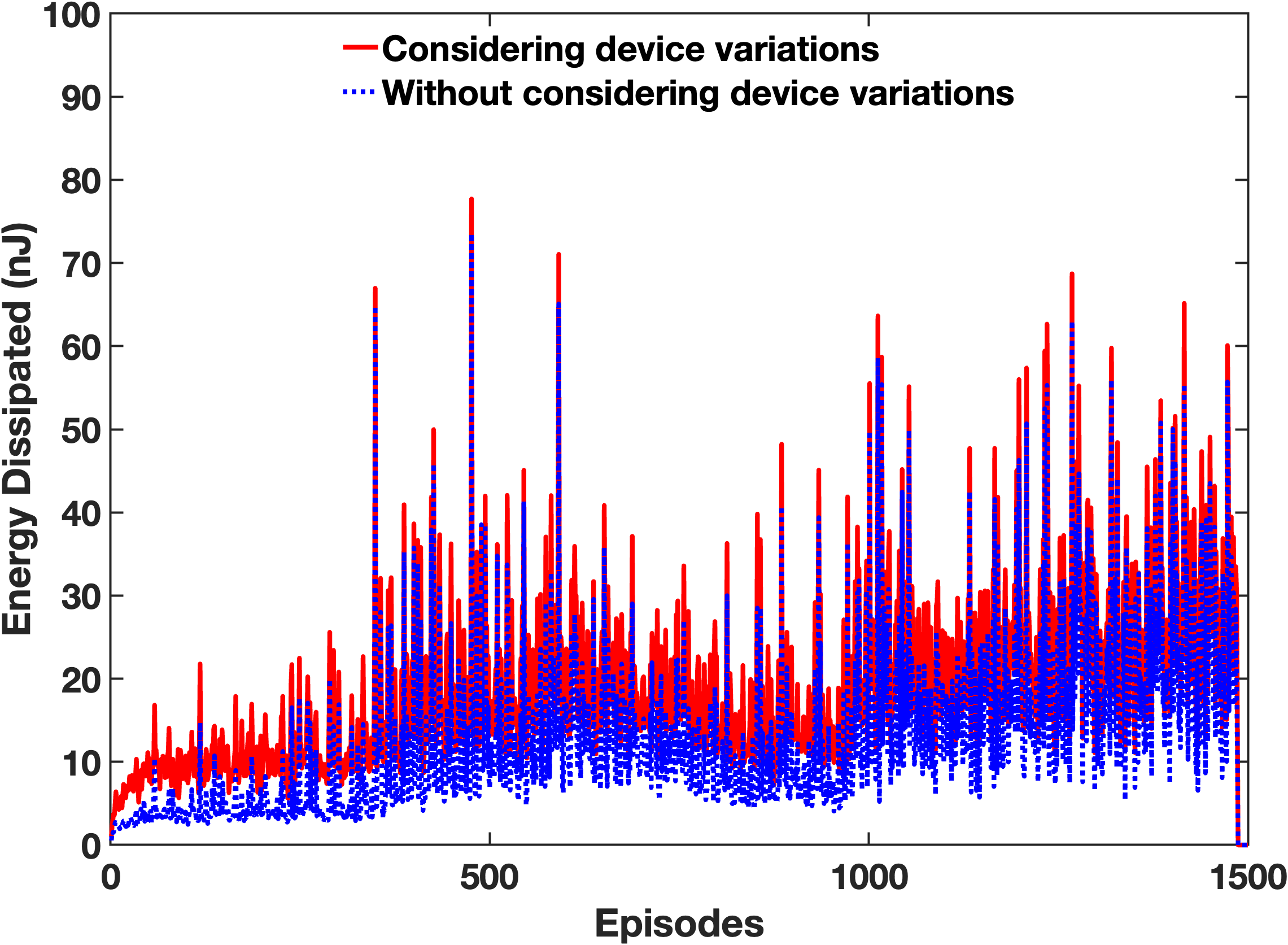}
    \caption{Energy consumed by the proposed implementation in each epoch of \textit{in situ} training.}
    \label{fig:energy_dissipated}
\end{figure}

\begin{figure}[t]
    \centering
    \includegraphics[width=0.5\textwidth]{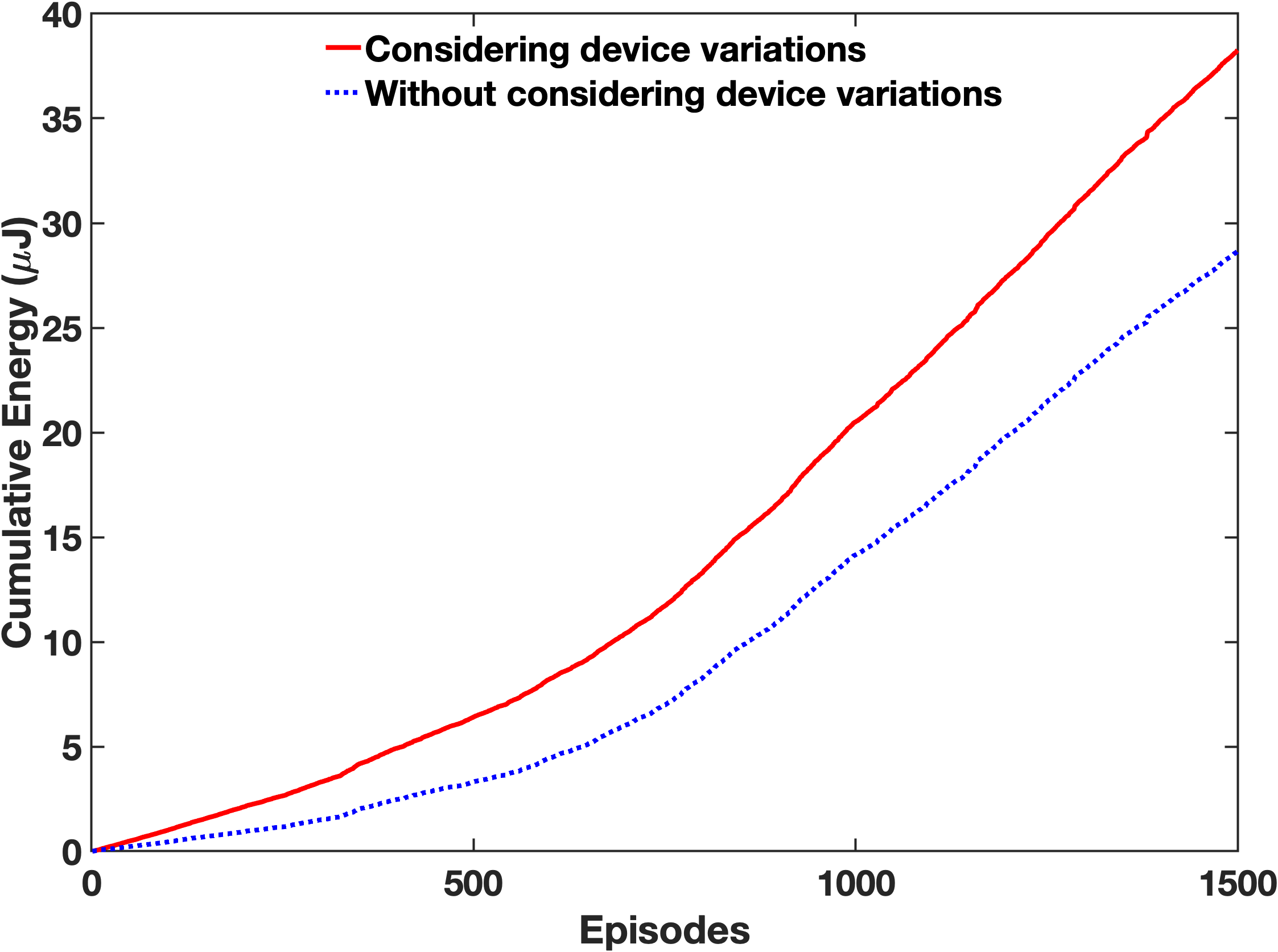}
    \caption{Cumulative energy consumed by the proposed implementation of Monte Carlo learning in each epoch of \textit{in situ} training.}
    \label{fig:cumulative_energy}
\end{figure}

\subsection{Energy Consumption}
The energy consumed during each episode of the \textit{in situ} training of proposed implementation considering nonidealities \cite{RRAMarray2,ManhattanRule} as well as noise-free passive RRAM crossbar is shown in Fig. \ref{fig:energy_dissipated}. While the cumulative energy consumed by the implementation on passive RRAM crossbar array considering noise and variability is 37.5 \si{\micro\joule}, the noise-free implementation consumes an energy of 28 \si{\micro\joule} as shown in Fig. \ref{fig:cumulative_energy}.

\subsection{Number of Program Cycles}
Due to the limited endurance of RRAMs ($\sim10^5$), the number of times each RRAM cell of the crossbar array can be switched during the \textit{in situ} training of proposed implementation becomes an operational bottleneck. Moreover, for the Q-learning-based RL algorithms, the agent explores and learns the best possible set of actions in a given environment utilizing a significantly large number of programming cycles \cite{1T1R_RL}. However, for the proposed implementation of Monte Carlo learning, we observed that the number of programming cycles for each RRAM cell is less than the endurance of the RRAMs. Since the proposed implementation creates a copy of the weight matrix at the beginning of each episode, the number of times each RRAM cell in the return matrix is programmed is much higher as compared to the weight matrix. However, even for the RRAMs in the return matrix, the maximum number of programming cycles is around $\sim10^3 - 10^4$ which is significantly less than the reported endurance of the practical RRAMs.



\subsection{Area Utilized}
The total area required to train an agent in Cart-Pole environment using Deep Q-Learning on active 1T-1R crossbar array is estimated as 12.23 \si{\milli\metre^2} \cite{1T1R_RL} whereas the proposed implementation of Monte Carlo learning on passive RRAM crossbar array with similar performance occupies an area of 103.68 \si{\micro\metre^2}. Therefore, the proposed implementation exhibits a footprint reduction by five orders of magnitude ($\sim1.18\times10^5$).

\section{Conclusion}
In this work, for the first time, we have performed an algorithm-hardware co-design for efficient implementation of reinforcement learning and proposed a novel and hardware-friendly RRAM endurance-compatible implementation of Monte Carlo learning on passive RRAM crossbar array with significantly low footprint and ultra-low power consumption. Our results indicate that the proposed implementation outperforms the prior RL implementations by more than five orders of magnitude in terms of area while exhibiting similar accuracy. These promising results may provide incentive for its experimental realization.

\bibliographystyle{ieeetr}
\bibliography{main.bib}

\end{document}